# Thermoelectric graphene photodetectors with sub-nanosecond response times at Terahertz frequencies

Leonardo Viti,[1] Alisson R. Cadore,[2] Xinxin Yang,[3] Andrei Vorobiev,[3] Jakob E. Muench,[2] Kenji Watanabe,[4] Takashi Taniguchi,[4] Jan Stake,[3] Andrea C. Ferrari,[2] Miriam S. Vitiello[1]

1. NEST, Istituto Nanoscienze - CNR and Scuola Normale Superiore, Piazza San Silvestro 12, 56127 Pisa, Italy
2. Cambridge Graphene Centre, University of Cambridge, 9, JJ Thomson Avenue, Cambridge CB3 0FA, UK
3. Department of Microtechnology and Nanoscience, Chalmers University of Technology, SE-41296 Gothenburg, Sweden
4. National Institute for Materials Science, 1-1 Namiki, Tsukuba, 305-0044, Japan

**Abstract**

Ultrafast and sensitive (noise equivalent power < 1 nWHz$^{-1/2}$) light-detection in the Terahertz (THz) frequency range (0.1-10 THz) and at room-temperature is key for applications such as time-resolved THz spectroscopy of gases, complex molecules and cold samples, imaging, metrology, ultra-high-speed data communications, coherent control of quantum systems, quantum optics and for capturing snapshots of ultrafast dynamics, in materials and devices, at the nanoscale. Here, we report room-temperature THz nano-receivers exploiting antenna-coupled graphene field effect transistors integrated with lithographically-patterned high-bandwidth (~100 GHz) chips, operating with a combination of high speed (hundreds ps response time) and high sensitivity (noise equivalent power ≤120 pWHz$^{-1/2}$) at 3.4 THz. Remarkably, this is achieved with various antenna and transistor architectures (single-gate, dual-gate), whose operation frequency can be extended over the whole 0.1-10 THz range, thus paving the way for the design of ultrafast graphene arrays in the far infrared, opening concrete perspective for targeting the aforementioned applications.

**Introduction**

Hot-carrier assisted photodetection is an efficient and inherently broadband detection mechanism in single layer graphene (SLG) [1-4]. When a photon is absorbed by the electronic population (either via interband or intraband transitions), the photoexcited carriers can relax energy through electron-electron scattering [5,6] or emission of optical phonons [5,6], on a time scale of 10 to hundreds fs [5,6]. However, the electron-to-lattice relaxation via acoustic phonons is slower (1-2 ps) [6], leading to a *quasi*-equilibrium state where the thermal energy is distributed amongst electrons [5,6] and not shared with the lattice. This produces an intriguing scenario, where the energy is absorbed by a system with an extremely low thermal capacitance ($c_e$ ~ 2000 $k_B$μm$^{-2}$, $k_B$ is the Boltzmann constant) [7-10], thus leading to the ultrafast (∼fs−ps) onset of thermal gradients in SLG-based nanostructures. At terahertz (THz) frequencies this effect is more relevant, since the emission of optical phonons is energetically forbidden [11], thus hindering this additional pathway for energy relaxation.

SLG is therefore a promising material for engineering high-speed (~ps response time) optoelectronic THz devices that could benefit from the above mechanism [12].

The detection of THz light is important for applications in imaging [13], tomography [14], security [15,16], biomedicine [17], and quantum optics [18]. An ideal THz photodetector (PD) should have a low noise equivalent power (NEP < nW/Hz$^{1/2}$), a large dynamic range (ideally >3 decades), have high detection speed (< ns), be broadband (0.1-10 THz), and operate at room temperature (RT). However, current RT THz PDs fail in targeting this combination of sensitivity, speed, and spectral range [19]. Graphene-based THz detectors relying on different physical mechanisms [4] have been widely demonstrated in the last few years [2,12,20-28] and include nanodevices exploiting the photovoltaic (PV) [22], the bolometric [23], the photothermoelectric (PTE) [2,12,27] and the plasma wave (PW) or Dyakonov-Shur effects, the latter in either its non-resonant [20,25] or resonant (at low temperatures) [26] configurations. At RT, PTE photodetectors have proven to be the most sensitive and fast [2,12,27], due to the occurrence of photo-induced temperature gradients which alter the electronic thermal distribution on a fast (~100 fs) timescale [5,6] and to the absence of an applied *dc* current through the SLG channel, which usually increases the noise level (dark current) in alternative configurations [23]. PTE detectors were demonstrated to reach response times ~100 ps at 1 THz [12]. The best combination of performance at frequency above 3 THz was achieved in a thermoelectric RT graphene device [2], showing simultaneously NEP < 100 pWHz$^{-1/2}$, response time $\tau$ ~40 ns (setup-limited), and a 3 orders of magnitude dynamic range. In this device, an *ad hoc* dual-gated, *H*-shaped antenna, having a strongly sub-wavelength gap (100 nm), defined a *p-n* junction, to which the performance improvement was ascribed. More recently, NEP≤160 pWHz$^{-1/2}$ with response times of 3.3 ns have been also reported in thermoelectric receivers exploiting broadband bow-tie antennas [27].

Here, we undertake the task of boosting detection performances with respect to that benchmark. We exploit two different architectures: a single-gated hBN/graphene/hBN field effect transistor (GFET) (Figure 1C) and a split-gate hBN/graphene/hBN *p-n* junction (Figure 1D). By deeply investigating the photodetection mechanism, we show that, independently from the geometry, both the architectures operate mainly via the PTE effect. We then evaluate and compare the detection performances, proving that $\tau$ can be lowered at the hundreds ps level, without spoiling the detector sensitivity. This is achieved as follows. First, we minimize the absorption area in the GFET channel. This allows maximizing the temperature increase within the electronic thermal distribution, since a smaller absorption area entails a smaller amount of carriers to be heated by the incoming electromagnetic field, and, in turn, a larger temperature increase [2]. Secondly, as a further refinement, we employ a novel electrodes design, which features on-chip transmission lines with bandwidth > 100 GHz, and readout electronics having bandwidth > 1 GHz.

By embedding the hBN/SLG/hBN layered materials heterostructures (LMH) [29,30] in FET coupled to on-chip planar THz antennas (Figure 1A,B), we demonstrate ultrafast ($\tau$ < 1 ns) detection of > 3 THz light at RT, with a record combination of speed, NEP and sensitivity, independent on architecture. This is possible owing to the fast (~100 fs) onset of thermal gradients along the SLG channel and the subsequent generation of a PTE photovoltage [1], not dependent on the selected architecture. Thus, encapsulated SLG-based devices coupled to antenna structures can be used for the characterization of high (> 10 MHz) repetition rate THz sources and high-speed (< 1 ns) and low noise (NEP < 1 nWHz$^{-1/2}$) THz imaging.

**Results and Discussion**

We engineer two photodetector configurations as follows. Sample A is an hBN encapsulated GFET integrated with a planar bow-tie antenna, asymmetrically connected to the source ($s$) and top-gate ($g_T$) electrodes, Figure 1C. Sample B is an hBN encapsulated GFET where two split-gates ($g_{TL}$, left gate and $g_{TR}$, right gate, Figure 1D), connected to the two branches of a linear dipole antenna, define a $p$-$n$ junction at its center [2]. Such antenna geometries are widely used in THz optoelectronics [2,4,24,31] and both enable broadband operation [2,32].

The hBN encapsulated GFET devices are fabricated as follows. hBN crystals are grown by the temperature-gradient method under high pressures and temperatures [33]. Bulk graphite is sourced from Graphenium. hBN and SLG are individually exfoliated on SiO$_2$/Si by micromechanical cleavage [34]. Initially, optical contrast [35] is utilized to identify SLG [29,30]. The transfer technique employs a stamp of polydimethylsiloxane (PDMS) and a film of polycarbonate (PC) mounted on a transparent glass slide for picking up the layered materials and transfer them to the final and undoped SiO$_2$/Si substrate. The presence and quality of SLG is then confirmed by Raman spectroscopy [36] (see *Methods*). The thickness of hBN is determined by atomic force microscope (AFM) and Raman spectroscopy [37,38]. Combining the results from optical microscopy, Raman spectroscopy and AFM, blister-free areas with full width at half maximum (FWHM) of the 2D peak, FWHM(2D) < 18 cm$^{-1}$ are selected for device fabrication.

Following their assembly, we process the heterostructures into antenna-coupled FETs. The GFET channel is first shaped by electron beam lithography (EBL), followed by dry etching of hBN and SLG [39] in SF$_6$. The SLG channel geometry is schematically represented in Figure 1: the channel is $L_C$ = 3 μm long and $W_C$ = 0.8 μm wide. The contact regions have lateral extensions. By simple geometrical considerations, it can be demonstrated that these extensions increase the perimeter of the stack, *i.e.* the length of the edge-contacts, thus reducing the contact resistance by 30%, with respect to a more standard rectangular channel geometry. Edge Au/Cr electrodes are defined by standard EBL [39,40], followed by metallization (40:5nm) and *lift-off*.

We employ, for both sample A and B, bottom hBN flakes of almost identical thickness ($h$), in order to make the comparison of the device performances consistent and reproducible. It is indeed worth mentioning that, due to the decrease of the electron-hole charge fluctuations at the substrate [41], changes of the bottom hBN layer thickness can significantly affect the FET mobility [29,42]. In the present case, the flakes thickness, retrieved by AFM are: bottom hBN $h$ = 23 nm, top hBN $h$ = 8 nm, for sample A, and bottom hBN $h$ = 25 nm, top hBN $h$ = 17 nm for sample B. The low thickness of the heterostructures (< 45 nm) and of the edge-contacts (~45 nm) allows us to use a thinner oxide (70 nm) as encapsulating layer before $g_T$ deposition (Figure 1C,D), thus increasing the effective gate-to-channel capacitance per unit area: $C_g$ ~ 100 nFcm$^{-2}$ for both samples. This parameter is important for THz FET detectors [25], since the responsivity ($R_v$), a figure of merit defined as the ratio between photovoltage ($\Delta u$) and impinging optical power, is typically proportional to the sensitivity of the FET conductance to changes in the gate voltage ($V_g$) [25].

In order to reduce parasitic capacitances, usually detrimental for high-speed (> 1 GHz) detection, and simultaneously minimize parasitic losses [43], we design and fabricate a microwave transmission line connected to the *s* and drain (*d*) edge-electrodes based on a coplanar strip-line (CPS) geometry [24], Figure 1B. We use this radio frequency (RF) on-chip component because of its simplicity. In contrast to the standard strip-line geometry [44], it does not require a ground plane, and, unlike the coplanar waveguide architecture [44], it consists of only two parallel metallic strips on the substrate top surface. In our devices, the strips are separated by a 2 μm gap, where one conductor (ground electrode, *s*) provides the electrical ground for the other (signal electrode, *d*). This architecture shows an almost perfect transmission below 30 GHz, with $S_{21} = 0$ dB, $S_{11} < -40$ dB, whereas at 3.4 THz the transmission is reduced, but not canceled, with $S_{21} = -3.5$ dB and $S_{11} = -25 \div -35$ dB (details about simulations are given in Supplementary Information). The transmission of the THz signal between the antenna-coupled GFET and the contacts can be detrimental for the overall detector performance. This is mainly due to the fact that the antenna modes lose energy (resulting in a decreased quality factor at resonance) if the antenna is not isolated from the surrounding circuit. Therefore, our design also includes a low-pass hammer-head filter along the CPS (Figure 1B) [45], with a cutoff frequency $f_{cut-off} \sim 300$ GHz, which enhances the isolation between antenna and readout circuit. It consists of a capacitive shunt with a lumped capacitance $C_f = 500$ aF. The dimensions of the structure are optimized by the simulation of the time-domain solver in CST Microwave Studio (see Supplementary Information).

The presence of the filter leaves the *S*-parameters almost unaltered for frequencies < 30 GHz: $S_{21} = 0$ dB, $S_{11} < -30$ dB. On the other hand, it modifies the transmission line properties at 3.4 THz: $S_{11} = -4$ dB, $S_{21} \sim -24$ dB. To further increase the signal extraction from the active element, the CPS has an adiabatically matched transition [46] between bonding pads and GFET electrodes, which hinders the formation of spurious reflections and consequent losses.

After this common protocol, samples A and B are processed following different architectures. For sample A, Figure 1C, the lobe of a THz planar bow-tie antenna (110 nm thick) is connected to the *s* electrode. Then, a thin top-gate oxide bi-layer is placed on the LMH, also covering the *s* and *d* contacts: 20 nm $HfO_2$ deposited via atomic layer deposition (ALD) and 50 nm $Al_2O_3$ deposited via Ar sputtering. The PD is finalized by the fabrication of $g_T$, in the shape of the arm of a bow-tie antenna, thus forming a complete bow-tie together with the *s* electrode. The antenna radius is 21 μm and the gap between antenna arms is 250 nm (Figure 1C). For sample B, Figure 1D, the same oxide bi-layer is deposited before the antenna fabrication. The antenna is here shaped as a linear dipole, with 24 μm arms separated by a gap of 90 nm (Figure 1D, further images are reported in the Supplementary Information). The two branches of the antenna also serve as top split-gates for the GFET. The gate voltages ($V_{gL}$, left gate bias and $V_{gR}$, right gate bias) can be individually controlled in order to create, at the center of the active channel, a *p-n* junction whose size is approximately corresponding to the gap between the two split-gates [2,47]. The gate geometry is therefore nominally the only difference between the two samples.

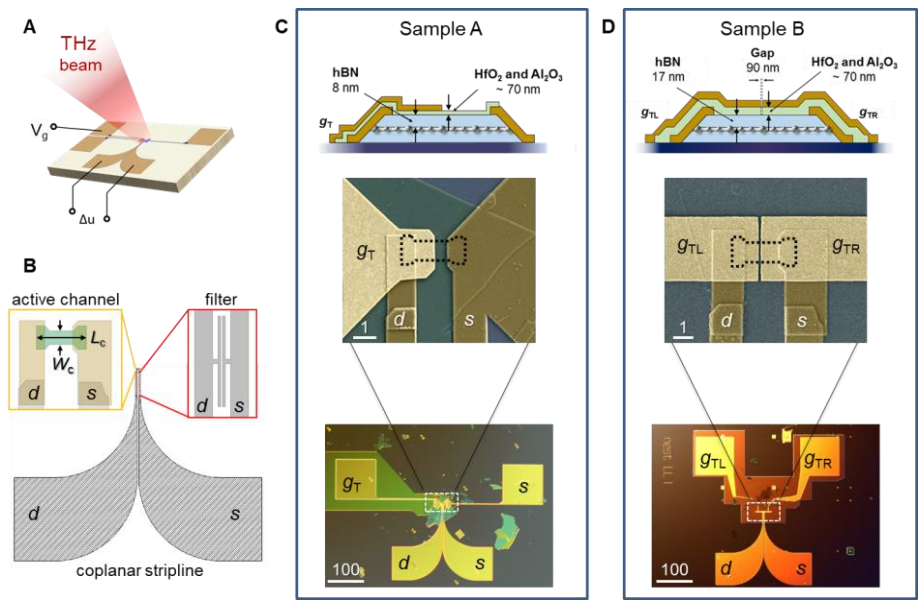

**Figure 1. Detector layout.** (A) Photodetector schematic: THz radiation is coupled to the GFET by a planar antenna and the photoresponse is recorded as a *dc* photovoltage (Δ*u*) between the *s* and *d* electrodes. (B) On-chip RF components. The *s* and *d* electrodes are shaped in a CPS geometry. Inset (left): the shape of the active LMH channel (green area) guarantees a lower contact resistance with respect to a rectangular geometry. The *s* and *d* contacts have a thickness of 45 nm in vicinity proximity of the GFET channel (yellow areas) and a thickness of 140 nm far from the GFET channel. Inset (right): planar low-pass filter, with cut-off frequency 300 GHz. (C) Sample A. Top: schematics of the LMH and electrodes layout, highlighting the different layer thicknesses. False color SEM image of the top-gated GFET (center) and optical microscope overview (bottom), where the bow-tie antenna position is marked with a dashed box. (D) Sample B. Top: schematics of the LMH and contacts design. False color SEM image of the GFET showing the split-top-gate geometry with the 90 nm gap (center) and optical microscope overview (bottom), where the position of the planar dipole antenna is marked with a dashed box. All scale bars are in units of micron.

The devices are then characterized electrically and optically at RT. The two-probe GFET transfer curve, measured for sample A in Figure 2A, shows a channel resistance (R) peak at $V_g = -4.6$ V (charge neutrality point, CNP). The extracted field-effect mobility ($\mu_{FE}$) is 17000 cm$^2$V$^{-1}$s$^{-1}$ for holes and 19000 cm$^2$V$^{-1}$s$^{-1}$ for electrons, with a residual carrier density $n_0 \sim 9\times10^{11}$ cm$^{-2}$. This is fitted using the formula [48] $R = R_0 + (L_C/W_C)\cdot(1/n_{2d}e\mu_{FE})$, where $R_0$ is the contact resistance and $n_{2d}$ is the gate-dependent charge density, given by [48] $n_{2d} = [n_0^2 + (C_g/e\,(V_g - V_{CNP}))^2]^{1/2}$.

We then test the room temperature sensitivity using a focused 3.4 THz beam with an average power $P_t = 100$ μW (see *Methods*). The intensity distribution on the focal plane (Figure 2D, sample A), displayed through the *xy* map of Δ*u*, unveils the Airy pattern [49] of the focused beam, showing four concentric rings (maxima) with the central Airy disk. This demonstrates the good signal-to-noise ratio (~1000 at $P_t = 100$ μW) of the proposed device. From the two-dimensional Gaussian fit of the intensity distribution in Figure 2D, we obtain standard deviations $\sigma_x = 95 \pm 1$ μm and $\sigma_y = 87 \pm 1$ μm along the *x* and *y* directions, respectively, from which we infer FWHM ~ 303 ± 2 μm (see Supplementary Information for further details). This is used to estimate the fraction of total power that impinges on the detector $P_a = P_t \cdot (A_\lambda/A_{spot}) = 2.7$ µW, where $A_\lambda = \lambda^2/4$ = $1.9 \times 10^{-3}$ mm$^2$ is the diffraction limited area (see Supplementary Information) and $A_{spot} = \pi \cdot (FWHM/2)^2$ = $72 \times 10^{-3}$ mm$^2$ is the beam spot area. Then, by measuring $\Delta u$ (see *Methods*) as a function of $V_g$ and dividing the as-obtained values by $P_a$, we retrieve the plot of $R_v$ as a function of $V_g$ (Figure 2B). The maximum $R_v = 30$ VW$^{-1}$ is obtained for $V_g = -7$V and the trend is compatible with a dominant PTE response (see Supplementary Information). This is corroborated by the following argument. At $V_{sd} = 0$V, in a single-gated GFET, connected by identical metallic layers at the *s* and *d* contacts, both the PTE and the non-resonant PW detection mechanisms can in principle be activated [25,27]. In the geometry of sample A, the PTE photovoltage reads $\Delta u_{PTE} = \Delta T_e \cdot (S_g - S_u)$ [25,27,31], where $\Delta T_e$ is the THz-induced electronic temperature difference between the (hot) source side of the channel, corresponding to the gap at the center of the bow-tie antenna, and the (cold) drain side (Figure 1C), $S_u$ is the Seebeck coefficient of the ungated region between the *s* and *g* electrodes and $S_g$ is the Seebeck coefficient of the gated LMH channel. By imposing $S_u = S_g$ for $V_g = 0$V and assuming $\Delta T_e$ weakly dependent on $V_g$ [2,25], we can analytically compute the gate voltage dependence of $\Delta u_{PTE} \propto S_g - S_u$ (see Supplementary Information for further details). The same argument applies to the overdamped PW photovoltage [20,25], $\Delta u_{PW} \propto -\sigma^{-1}(\partial\sigma/\partial V_g)$. The comparison between $\Delta u_{PTE}(V_g)$, $\Delta u_{PW}(V_g)$ and the experimental $R_v(V_g)$ curves (Figure 2B) unveils that the PTE effect well matches with our experimental observation and better reproduces our data with respect to the PW model, which predicts that the maximum response (in absolute value) occurs at $V_g = -3.5$V and the responsivity is finite and negative for $V_g = 0$V, in stark contrast with our measurements, where $R_v \approx 0$ VW$^{-1}$ at $V_g = 0$V.

This conclusion is further supported by the temperature (*T*) dependent analysis of the responsivity, which unambiguously shed light on the core detection dynamics. To this purpose we mount the detector in a He flux cryostat and we vary the heat sink *T* in the 6 – 260 K range. The measured responsivity (Figure 2E) shows a non-monotonic behavior as a function of *T*, with a maximum around a crossover temperature $T^* = 60$ K, in agreement with what observed in other spectral ranges [50]. The origin of such a behavior can be retrieved by the analysis of the electron cooling dynamics in SLG. $\Delta u_{PTE}$ is proportional to $\Delta T_e$, which, in turn, is proportional to the cooling length $\xi = (k/\gamma c_e)^{1/2}$ [1,2,50] (the proportionality holds as long as $\xi < L_C$), where $k$ is the thermal conductivity and $\gamma$ is the cooling rate. Since both $k$ and $c_e$ scale linearly with *T*, the functional dependency of the cooling length $\xi$ (and $\Delta u_{PTE}$), with respect to *T*, is the same as $\gamma^{-1/2}$. For $T<T^*$, $\gamma(T)$ is dominated by acoustic phonon emission and scales as $\sim T^{-1}$, whereas at higher *T*, the disorder-assisted scattering (*supercollision*) gives rise to a competing cooling channel which follows the power law $\gamma \sim T$ [50]. The two effects give rise to a crossover temperature ($T^*$) for which $\gamma$ is minimum and, consequently, $\Delta u_{PTE}$ is maximum. We then compare the temperature dependence of $R_v$ at two distinctive gate voltages, $V_g = -5$V (close to CNP, low carrier density, $n_{2d} \sim 10^{12}$ cm$^{-2}$) and at $V_g = -9$ V (away from CNP, holes density up to $n_{2d} \sim 4 \times 10^{12}$ cm$^{-2}$). The non-monotonic behavior is more evident at lower $n_{2d}$, in qualitative agreement with previous findings on PTE detection [25,50]. In a non-degenerate electron system, $\Delta u_{PTE}(T)$ is completely determined by $\Delta T_e$, being the Seebeck coefficient weakly dependent from *T* [25]; conversely, in the degenerate case, *S* is proportional to *T* [51] and compensates the decrease of $\Delta T_e$ at higher *T*, resulting in an almost *T*-independent $\Delta u_{PTE}$. For sample

A, under the assumption of a *noise spectral density* (NSD, *i.e.* noise power per unit bandwidth) dominated by thermal fluctuations [31] (see Supplementary Information), we estimate NEP = $1/R_v \cdot (4k_B RT)^{1/2}$. The NEP curve as a function of $V_g$ (Figure 2C) shows a minimum NEP ~350 pWHz$^{-1/2}$ at $V_g = -7$ V.

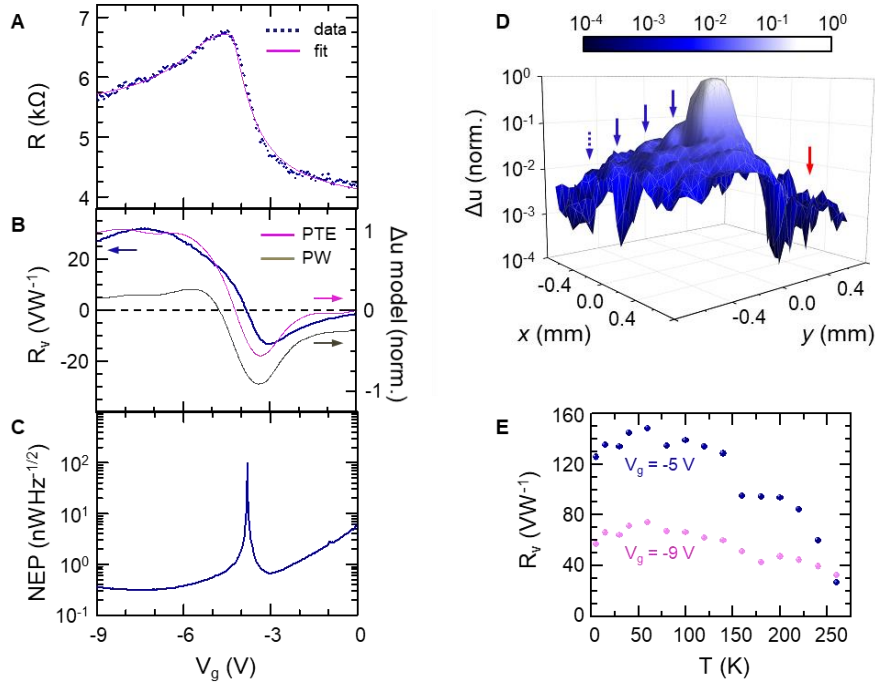

**Figure 2**. **Electrical and optical characteristics of single-gate GFET.** (A) Electrical resistance $R$ as a function of $V_g$ at RT in a two-terminal configuration. (B) $R_v$ at RT as a function of $V_g$ (left vertical axis), compared with the normalized expected photothermoelectric and overdamped plasma wave photovoltages (right vertical axis). (C) NEP calculated as a function of $V_g$ under the assumption of Johnson-Nyquist dominated noise spectral density [2]. A minimum NEP~350 pW Hz$^{-1/2}$ is obtained for $V_g = -7$V. (D) Logarithmic plot of the normalized photovoltage on the focal plane, for an average impinging THz power of 100 µW. The four Airy maxima are indicated by blue arrows on the left of the central Airy disk. The red arrow indicates a portion of the focal plane where the beam is blocked by the output window of the cryostat where the QCL is located. The FWHM of the beam is 303 µm. (E) $R_v$ plotted as a function of $T$ measured at $V_g = -5$V (blue dots) and $V_g = -9$V (magenta dots).

We use a similar approach for the optical and electrical characterization of sample B. Figure 3 plots the device performance as a function of bias applied at the split-gates. By independently varying the two gate voltages, we control the Fermi level ($E_F$) and, consequently, $n_{2d}$ on each side of the dual-gated SLG junction [2,47]. The color plot of $R$ with respect to $V_{gR}$ (right gate, horizontal axis) and $V_{gL}$ (left gate, vertical axis) in Figure 3A allows us to extract a hole and electron $\mu_{FE}$ ~19000 cm$^2$V$^{-1}$s$^{-1}$ and 15000 cm$^2$V$^{-1}$s$^{-1}$, respectively, with a residual carrier density $n_0$ ~1×10$^{12}$ cm$^{-2}$.

The independent control of the $E_F$ on each side of the junction allows individual control of the two Seebeck coefficients $S_L$ and $S_R$ [2,47], which can be used to maximize the photoresponse. THz detection in a graphene *p-n* junction is expected to be dominated by the PTE effect [2]. $\Delta u_{PTE}$, measured between the drain and source electrodes, can be written as [52]:

$$\Delta u_{PTE} = \int_D^S \frac{\partial T_e}{\partial x} \cdot S(x) dx = \Delta T_e \cdot (S_L - S_R) \qquad (1)$$

where $\Delta T_e$ is the electronic temperature increase a consequence of the absorption of THz radiation at the junction.

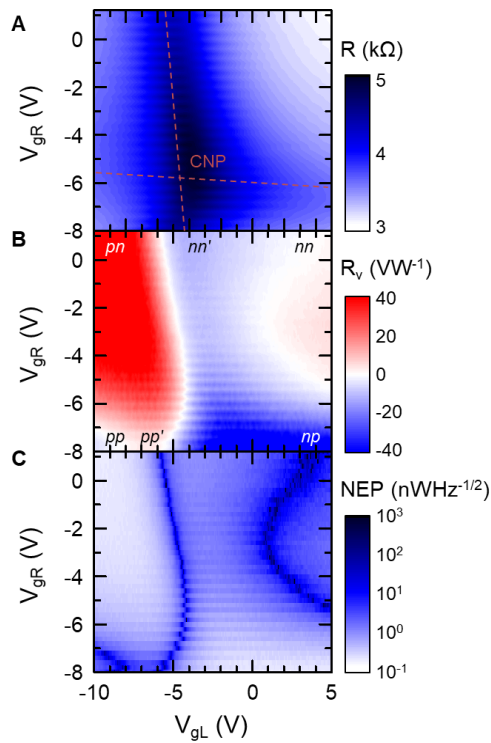

**Figure 3. Electrical and optical characteristics of double-gated graphene p-n junction.** (A) Analysis of electrical transport of GFET: two-terminal RT resistance as a function of split-gate bias. The dashed lines indicate the CNP positions for $V_{gL}$ and $V_{gR}$. (B) Color map of $R_v$ as a function of $V_{gL}$ and $V_{gR}$. $R_v$ undergoes many sign changes, corresponding to transitions between the different configurations of the *p-n* junction, attainable by polarizing the gates. (C) Two-dimensional plot of the NEP (logarithmic scale) as a function of $V_{gL}$ and $V_{gR}$. A minimum NEP of 120 pWHz$^{-½}$ is obtained for $V_{gL} = -8$ V and $V_{gR} = -4$ V.

Figure 3B is a color map of $R_v$ obtained by continuously changing $V_{gR}$ and $V_{gL}$ in the same ranges of Figure 3A. The maxima of $R_v$ (~50 VW$^{-1}$) are obtained when the two local gates have opposite polarity with respect to the CNP, *i.e.* in bi-polar junction configuration (*p-n, n-p*). The resulting *six-fold* pattern in the measured photovoltage is ascribed to the non-monotonic gate voltage dependence of $S_L$ and $S_R$ on each side of the junction, and is a unique fingerprint of a dominant hot-carrier assisted PTE effect in SLG [1,2,53]. Therefore, for the *p-n* junction, the room temperature $R_v$ characterization alone is sufficient to unambiguously unveil the dominant PTE THz detection.

From $R$ and $R_v$, we can estimate NEP of sample B, assuming thermal-noise limited operation. The contour plot of NEP as a function of the two gate voltages (Figure 3C) shows a minimum NEP ~120 pW Hz$^{-½}$ at $V_{gL} = -8$V and $V_{gR} = -4$V. Sample B is therefore ~3 times more sensitive than sample A. This can be attributed to the larger field enhancement provided by the dual-gate configuration, in particular to the narrow (90 nm) gap between the antenna arms, in agreement with Ref. [2].

To extract the response time and the bandwidth $BW = (2\pi\tau)^{-1}$, we shine light from a pulsed THz quantum cascade laser (QCL, pulse width ~150 ns and repetition rate 333 Hz) and record the signal with a fast oscilloscope (5 GS/s) after a pre-amplification stage (low noise voltage preamplifier, model Femto-DUPVA, bandwidth 1.2 GHz).

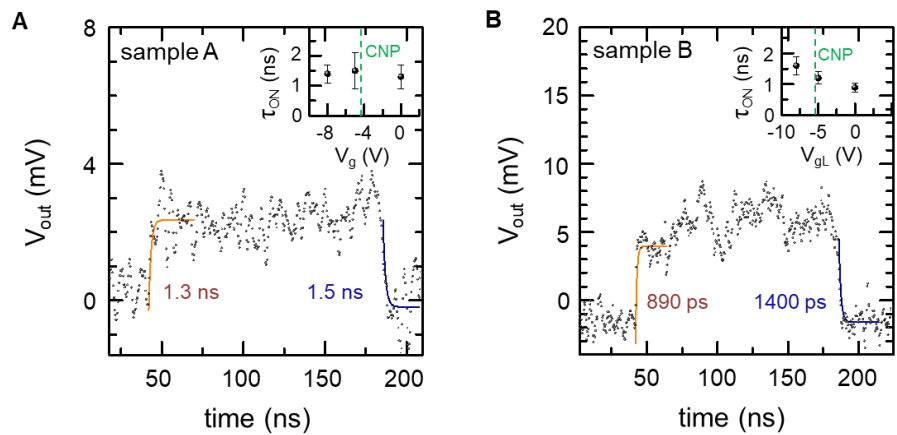

**Figure 4. Electrical bandwidth and response time.** (A) Photovoltage time-trace under illumination with a 150 ns THz pulse with a peak power of 10 mW recorded with sample A at $V_g$ = 0V. The time constants $\tau_{ON}$ = 1.3 ± 0.4 ns and $\tau_{OFF}$ = 1.5 ± 0.6 ns are obtained by fitting the data. Inset: variation of $\tau_{ON}$ as a function of $V_g$. The rise-time does not depend on the device resistance. (B) Time trace recorded with sample B at $V_{gL}$ = $V_{gR}$ = 0V, giving $\tau_{ON}$ = 890 ± 150 ps and $\tau_{OFF}$ = 1400 ± 250 ps. Inset: variation of $\tau_{ON}$ as a function of $V_g$. The rise-time does not depend on the device resistance.

Figures 4A,B show the time traces of samples A, B, recorded at zero gate bias with an oscilloscope having a temporal resolution ~200 ps. We extract the rise-time $\tau_{ON}$ and fall-time $\tau_{OFF}$ by using the fitting functions $V_{out} = c_0 + V_{ON} \cdot [1-\exp(-(t-c_1)/\tau_{ON})]$ and $V_{out} = c_2 + V_{OFF} \cdot \exp(-(t-c_3)/\tau_{OFF})$, where $c_0$, $c_1$, $c_2$, $c_3$ are fitting parameters, and $V_{ON}$ and $V_{OFF}$ are the voltage jumps in the waveforms corresponding to the rising-edge and falling-edge. We find similar results for both devices, with rise-times slightly shorter with respect to fall-times. Sample A shows $\tau_{ON}$ = 1.3 ± 0.4 ns and $\tau_{OFF}$ = 1.5 ± 0.6 ns at $V_g$ = 0V, sample B shows $\tau_{ON}$ = 890 ± 150 ps and $\tau_{OFF}$ = 1.4 ns ± 0.25 ns at $V_{gL}$ = $V_{gR}$ = 0V. These response times are, to the best of our knowledge, the lowest in GFET devices with NEP < 1 nWHz$^{-½}$. In terms of $BW$, considering the lower values of $\tau$ as limit response time, we obtain $BW$ = 125 ± 35 MHz for sample A and $BW$ = 180 ± 30 MHz for sample B, *i.e.* 50 times better than in Ref. [2]. The small discrepancy between the latter values can be ascribed to fluctuations in the QCL output power, possibly caused by time jitter (± 100 ps [54]) in the electrical circuit employed to drive the laser.

To further validate this assessment, we measure the detector rise-time under different configuration of gate voltages, *i.e.* at different charge density and SLG resistance. The response time of a PD is ultimately limited by the *RC* time constant of the circuit [2]. Therefore, if the PD is the key element limiting the detection speed, a change in *R* should directly and proportionally reflect into a change in $\tau$, via $\tau = R \cdot C$. We thus select and investigate three gate voltage configurations for both devices. The results are shown in the insets of Figures 4A,B.

For sample A, we obtain $\tau_{ON} = 1.3 \pm 0.4$ ns at $V_g = 0$ V ($R = 4.2$ k$\Omega$), $\tau_{ON} = 1.5 \pm 0.6$ ns at $V_g = -5$V ($R = 6.4$ k$\Omega$) and $\tau_{ON} = 1.4 \pm 0.3$ ns at $V_g = -8$V ($R = 5.7$ k$\Omega$), showing the lack of a direct proportionality relation between $R$ and $\tau_{ON}$. The same conclusion can be drawn for sample B at $V_{gR} = 0$V, where $\tau_{ON} = 890 \pm 150$ ns for $V_{gL} = 0$V ($R = 3.7$ k$\Omega$), $\tau_{ON} = 1.2 \pm 0.2$ ns for $V_{gL} = -5$V ($R = 4.7$ k$\Omega$), and $\tau_{ON} = 1.6 \pm 0.3$ ns for $V_{gL} = -8$V ($R = 4.0$ k$\Omega$). This demonstrates that $\tau$ is not affected by the SLG resistance in the tested range. This illustrates that the PD itself is not limiting the measured maximum speed, which is instead affected by the switching time of the QCL. A higher intrinsic speed beyond the set-up limited value is in good agreement with reports of high-speed, PTE-based SLG detectors for integrated photonics, with reported 3 dB BW in the tens of GHz [47]. In this work, high-speed performance is enabled by the on-chip architecture, featuring RF electronic components, which mitigates the presence of parasitic capacitances and the undesirable crosstalk between sensing element and outer on-chip components.

Our results show that, up to a bandwidth of 150 MHz, the two proposed architectures are substantially equivalent. Both configurations lead to $\tau \sim$ ns, even though the two geometries are different: in sample A the THz field is distributed along the un-gated portion of the channel (250 nm), whereas in sample B the two symmetric split gates, defining a narrow gap (90 nm), provide a more localized enhancement of the THz field at the center of the SLG channel. The speed limit is, in both cases, lower than that reported in Ref. [2], the switching speed being limited by the onset speed and jitter noise of the employed QCL system. This equivalence is not surprising. As revealed by the low temperature characterization of sample A (Figure 2E), both architectures mainly operate through the same detection mechanism: the PTE effect. This is known to be the dominant mechanism for devices operating through *p-n* junction rectification [1,2], however it has also been observed in antenna-coupled single-gated architectures [20,25,26], where the antenna provided asymmetric THz excitation, essential for the activation of the PTE mechanism. Moreover, our data show that the speed of the two devices does not even depend on the existence of a *p-n* junction, but it only requires that the gates create an imbalance in the Seebeck coefficient along the graphene channel.

**Conclusions**

In summary, the performance achieved at RT on both devices demonstrates that PTE THz detectors, coupled with high-bandwidth on-chip (~ 100 GHz) and external electronics, detect pulses with sub-ns temporal extension, opening unique perspectives for ultrafast applications in a plethora of research field as ultrafast nano-spectroscopy, quantum science, coherent control of quantum nanosystems and high speed communications. Further improvements on the detection performances can be achieved via the on-chip integration of coplanar waveguides and pre-amplification stages. It is worth mentioning that, measuring the intrinsic speed limit of the PTE mechanism in SLG devices, which is expected to be $\tau \sim 10$ ps [2], would require completely avoiding the limitations set by the readout electronics. This could be obtained, for example, by exploiting interferometric techniques, such as pulse autocorrelation measurements [55].

Our results open a route for characterization of high repetition rate THz sources, transient effects in nonlinear optoelectronic devices (*e.g.* saturable absorbers), time-resolved intracavity-mode dynamics of THz QCL frequency combs and ultimately for high-speed and low noise THz imaging, never pioneered so far.

**Methods**

**Sample Characterization**

Raman measurements are performed using a Renishaw InVia spectrometer equipped with a 100× objective, 2400mm$^{-1}$ grating at 514 nm. The power on the sample is <1 mW to avoid any heating and damage. AFM is performed in tapping mode to characterize the topography and thickness of the LMHs using a Bruker Dimension Icon system. Figure 5A plots the spectra of a typical LMH, with 8 nm and 23 nm thickness top and bottom hBN flakes, while Figure 5B is a false color optical image of the LMH, highlighting the SLG edges. Figure 5A shows that the $E_{2g}$ peak for both bottom and top hBN are ~1366 cm$^{-1}$, with FWHM($E_{2g}$) ~9.3 and 9.7 cm$^{-1}$, consistent with bulk hBN [37]. Figure 5A plots the SLG G and 2D peaks before and after stacking. Before encapsulation, the 2D and G peaks have FWHM(2D)~27cm$^{-1}$, Pos(2D)~2682cm$^{-1}$, Pos(G)~1589cm$^{-1}$, FWHM(G)~8 cm$^{-1}$, and the intensity and areas ratio of 2D and G peaks are I(2D)/I(G)~1.4, A(2D)/A(G)~4.6, as expected for SLG with $E_F \geq 250$ meV [56,57]. No D peak is observed, indicating negligible defects [58]. After LMH assembling, the combined hBN $E_{2g}$ peak is at Pos($E_{2g}$)~1366 cm$^{-1}$, with FWHM($E_{2g}$)~9.5 cm$^{-1}$. For the encapsulated SLG we have Pos(2D)~2697 cm$^{-1}$, FWHM(2D)~17 cm$^{-1}$, Pos(G) ~1584cm$^{-1}$, FWHM(G)~14 cm$^{-1}$, I(2D)/I(G)~13, and A(2D)/A(G)~12, indicating $E_F \ll 100$meV [56,57]. The changes in FWHM(2D) after encapsulation indicates a reduction in the nanometer-scale strain variations within the sample [29,59]. Figure 5C shows a FHWM(2D) map across a bubble-free LMH sample, exhibiting homogeneous (spread < 1 cm$^{-1}$) and narrow (~17 cm$^{-1}$) FWHM(2D), which is selected for the GFET fabrication.

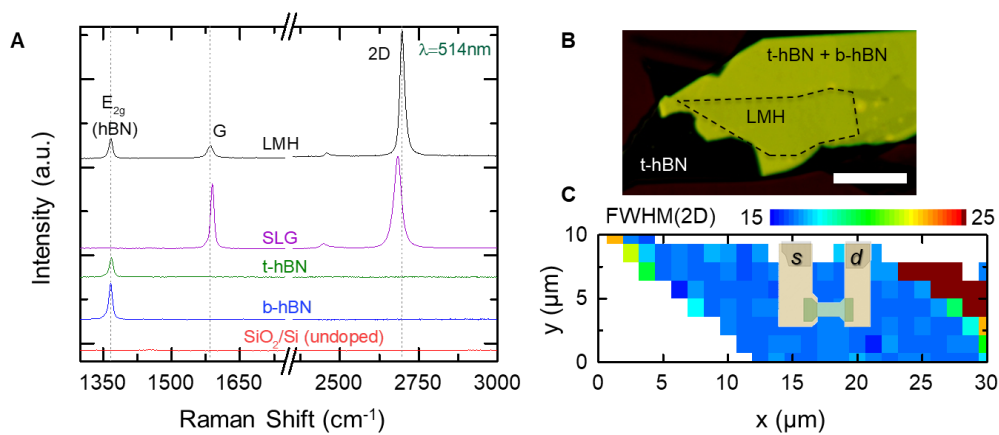

**Figure 5. Sample characterization and selection of the device area**. (A) Raman spectra before and after LMH assembly measured at 514nm. The bottom hBN (b-hBN) is shown in blue, the top one (t-hBN) in green, the SLG in purple, and the assembled LMH in black, while the SiO$_2$/Si substrate in red. The hBN $E_{2g}$, G, and 2D peaks are highlighted by the dashed grey lines. (B) False color optical image of the LMH. SLG is indicated by a black dashed line. Scale bar is 10 μm. (C) Spatial map of FWHM(2D), indicating the area which the GFET is designed.

**Optical measurements**

In order to test the PD sensitivity, we use a 3.4 THz quantum cascade laser (QCL), operating in pulse mode with a repetition rate of 40 kHz and a pulse width of 1 μs and refrigerated at 30 K by means of a Stirling cryocooler (estimated lattice temperature of the active region 170 K [60]). The divergent beam (divergence angle ~ 30°) is collimated and then focused using two picarin (tsupurica) lenses with focal lengths 50 and 30 mm. The average output power can be continuously varied up to ~1 mW at the PD position. The measurements are performed by keeping the *s* electrode grounded and by extracting the photovoltage signal Δ*u* at the *d* contact. The latter signal is then pre-amplified with a voltage pre-amplifier (FEMTO, input impedance 1 MΩ, gain 40 dB, *BW* 200 MHz) and recorded with a lock-in technique, referenced by a 1333 kHz square wave. Δ*u* is estimated as 2.2· $V_{LI}/\eta$ [31], where $V_{LI}$ is the lock-in signal $\eta$ is the voltage preamplifier gain coefficient. The detectors are mounted on a *xyz* stage allowing automated spatial positioning.

**Supplemental material**

The Supplementary Information file is available free of charge. Simulations of the on-chip RF filter; device layout, optical images; THz intensity distribution; Photothermoelectric effect; Responsivity evaluation; Noise equivalent power evaluation; Noise Equivalent Energy.

This is the authors' version of the submitted article, published in its final form at http://dx.doi.org/10.1515/nanoph-2020-0255

**Acknowledgements**

We acknowledge funding from the ERC Consolidator Grant SPRINT (681379) and the EU Graphene Flagship, ERC grants Hetero2D, GSYNCOR, EPSRC grants EP/L016087/1, EP/K01711X/1, EP/K017144/1. M.S.V. acknowledges partial support from the second half of the Balzan Prize 2016 in applied photonics delivered to Federico Capasso.

**Author contributions:** L.V and M.S.V. conceived the core idea. A.R.C and A.C.F. prepared the hBN/graphene/hBN heterostrcutures and characterized the quality of the graphene; K.W. and T.T. provided high quality hBN; X.Y., A.V. and J. S. contributed to the design of the microstrip line; L.V. fabricated the sample and performed electrical and optical measurements; L.V. and M.S.V analyzed the data and wrote the manuscript. All authors discussed the results and contributed to the writing of the manuscript. M.S.V. supervised the study.

**Competing financial interests:** The authors declare no competing financial interests.

**Data availability:** The data that support the plots within this paper and other findings of this study are available from the corresponding authors upon reasonable request.

**References**

[1] Gabor NM, Song JCW, Ma Q, et al. Hot carrier-assisted intrinsic photoresponse in graphene. Science 2011, 334, 648–652.
[2] Castilla S, Terrés B, Autore M, et al. Fast and Sensitive Terahertz Detection Using an Antenna-Integrated Graphene pn Junction. Nano Lett. 2019, 19, 2765-2773.
[3] Tielrooij KJ, Piatowski L, Massicotte M, et al. Generation of photovoltage in graphene on a femtosecond timescale through efficient carrier heating. Nat. Nanotechnol. 2015, 10, 437-443.
[4] Koppens FHL, Mueller T, Avouris Ph, Ferrari AC, Vitiello MS, Polini M. Photodetectors based on graphene, other two-dimensional materials and hybrid systems. Nat. Nanotechnol. 2014, 9, 780-793.
[5] Tomadin A, Brida D, Cerullo G, Ferrari AC, Polini M. Nonequilibrium dynamics of photoexcited electrons in graphene: collinear scattering, Auger processes, and the impact of screening. Phys. Rev. B 2013, 88, 035430.
[6] Brida D, Tomadin A, Manzoni C, et al. Ultrafast collinear scattering and carrier multiplication in graphene. Nat. Commun. 2013, 4, 1987.
[7] Fong KC, Schwab KC. Ultrasensitive and wide-bandwidth thermal measurements of graphene at low temperatures. Phys. Rev. X 2012, 2, 1-8.
[8] Pop E, Varshney V, Roy AK. Thermal properties of graphene: Fundamentals and applications. MRS Bull. 2012, 37, 1273–1281.
[9] Soavi G, Wang G, Rostami H, et al. Broadband, electrically tunable third-harmonic generation in graphene. Nat. Nanotechnol. 2018, 13, 583-588.
[10] Soavi G, Wang G, Rostami H, et al. Hot electrons modulation of third-harmonic generation in graphene. ACS Photonics 2019, 6, 2841-2849.
[11] Lazzeri M, Piscanec S, Mauri F, Ferrari AC, Robertson J. Phonon linewidths and electron-phonon coupling in graphite and nanotubes. Phys. Rev. B 2006, 73, 155426.
[12] Cai X, Sushkov AB, Suess RJ, et al. Sensitive room-temperature terahertz detection via the photothermoelectric effect in graphene. Nat. Nanotechnol. 2014, 9, 814-819.
[13] Chan WL, Deibel J, Mittleman DM, Imaging with terahertz radiation, Reports Prog. Phys. 2007, 70, 1325–1379.

[14] Guillet JP, Recur B, Frederique L, et al. Review of terahertz tomography techniques. J. Infrared, Millimeter, Terahertz Waves 2014, 35, 382–411.

[15] Kawase K, Ogawa Y, Watanabe Y, Inoue H. Non-destructive terahertz imaging of illicit drugs using spectral fingerprints. Opt. Express 2003, 11, 2549.

[16] Tonouchi M. Cutting-edge terahertz technology. Nat. Photon. 2007, 1, 97–105.

[17] Kašalynas I, Venckevičius R, Minkevičius L, et al. Spectroscopic terahertz imaging at room temperature employing microbolometer terahertz sensors and its application to the study of carcinoma tissues. Sensors 2016, 16, 1–15.

[18] Dhillon SS, Vitiello MS, Linfield EH, et al. The 2017 terahertz science and technology roadmap. J. Phys. D: Appl. Phys. 2017, 50, 043001.

[19] Sizov F. Terahertz radiation detectors: the state-of-the-art. Semicond. Sci. Technol. 2018, 33, 123001.

[20] Vicarelli L, Vitiello MS, Coquillat D, et al. Graphene field-effect transistors as room-temperature terahertz detectors. Nat. Mater. 2012, 11, 865-871.

[21] Mittendorff M, Winnerl S, Kamann J, et al. Ultrafast graphene-based broadband THz detector. Appl. Phys. Lett. 2013, 103, 021113.

[22] Degl'Innocenti R, Xiao L, Jessop DS, et al. Fast Room-Temperature Detection of Terahertz Quantum Cascade Lasers with Graphene-Loaded Bow-Tie Plasmonic Antenna Arrays. ACS Photonics 2016, 3, 1747-1753.

[23] Muraviev AV, Rumyantsev SL, Liu G, Balandin AA, Knap W, Shur MS. Plasmonic and bolometric terahertz detection by graphene field-effect transistor. Appl. Phys. Lett. 2013, 103, 181114.

[24] Generalov AA, Andersson MA, Yang X, Vorobiev A, Stake J. A heterodyne graphene FET detector at 400 GHz. 2017 42nd Int. Conf. Infrared, Millimeter, Terahertz Waves 2017, 1–2.

[25] Bandurin DA, Gayduchenko I, Cao I, et al. Dual origin of room temperature sub-terahertz photoresponse in graphene field effect transistors. Appl. Phys. Lett. 2018, 112, 141101.

[26] Bandurin DA, Svintsov D, Gayduchenko I, et al. Resonant terahertz detection using graphene plasmons. Nat. Commun. 2018, 9, 5392.

[27] Viti L, Purdie DG, Lombardo A, Ferrari AC, Vitiello MS. HBN-encapsulated, graphene-based, room-temperature terahertz receivers, with High Speed and Low Noise. Nano Lett. 2020, 20, 3169–3177.

[28] Auton G, But DB, Zhang J, et al. Terahertz Detection and Imaging Using Graphene Ballistic Rectifiers. Nano Lett. 2017, 17, 7015-7020.

[29] Purdie DG, Pugno NM, Taniguchi T, Watanabe K, Ferrari AC, Lombardo A. Cleaning interfaces in layered materials heterostructures. Nat. Commun. 2018, 9, 5387.

[30] De Fazio D, Purdie DG, Ott AK, et al. High-Mobility, Wet-Transferred Graphene Grown by Chemical Vapor Deposition. ACS Nano 2019, 13, 8926-8935.

[31] Viti L, Politano A, Zhang K, Vitiello MS. Thermoelectric terahertz photodetectors based on selenium-doped black phosphorus flakes. Nanoscale 2019, 11, 1995–2002.

[32] Coquillat D, Marczewski J, Kopyt P, Dyakonova N, Giffard B, Knap W. Improvement of terahertz field effect transistor detectors by substrate thinning and radiation losses reduction. Optics Express 2016, 24, 272-281.

[33] Watanabe K, Taniguchi T, Kanda H. Direct-bandgap properties and evidence for ultraviolet lasing of hexagonal boron nitride single crystal. Nat. Mater. 2004, 3, 404–409.

[34] Novoselov KS, Jiang D, Schedin F, et al. Two-dimensional atomic crystals. Proc. Natl. Acad. Sci. U. S. A. 2005, 102, 10451–10453.

[35] Casiraghi C, Hartschuh A, Lidorikis E, et al. Rayleigh imaging of graphene and graphene layers. Nano Lett. 2007, 7, 2711–2717.

[36] Ferrari AC, Meyer JC, Scardaci V, et al. Raman spectrum of graphene and graphene layers. Phys. Rev. Lett. 2006, 97, 187401.

[37] Reich S, Ferrari AC, Arenal R, Loiseau A, Bello I, Robertson J. Resonant Raman scattering in cubic and hexagonal boron nitride. Phys. Rev. B 2005, 71, 205201.

[38] Arenal R, Ferrari AC, Reich S, et al. Raman spectroscopy of single-wall boron nitride nanotubes. Nano Lett. 2006, 6, 1812-1816.

[39] Jessen BS, Gammelgaard L, Thomsen MR, et al. Lithographic band structure engineering of graphene. Nat. Nanotechnol. 2019, 14, 340-346.

[40] Wang L, Meric I, Huang PY, et al. One-Dimensional Electrical Contact to a Two-Dimensional Material. Science 2013, 342, 614-617.

[41] Xue J, Sanchez-Yamanishi J, Bulmash D, et al. Scanning tunnelling microscopy and spectroscopy of

ultra-flat graphene on hexagonal boron nitride. Nat. Mater. 2011, 10, 282–285.

[42] Dean CR, Young AF, Meric I, et al. Boron nitride substrates for high-quality graphene electronics. Nat. Nanotechnol. 2010, 5, 722–726.

[43] Kopyt P, Salski B, Marczewski J, Zagrajek P, Lusakowski J. Parasitic effects affecting responsivity of sub-THz radiation detector built of a MOSFET. J. Infrared, Millimeter, Terahertz Waves 2015, 36, 1059–1075.

[44] Lee TH. Planar Microwave Engineering. Cambridge University Press, Cambridge, 2004.

[45] Wang C, He Y, Lu B, et al. Robust sub-harmonic mixer at 340 GHz using intrinsic resonances of hammer-head filter and improved diode model. J. Infrared, Millimeter, Terahertz Waves 2017, 38, 1397–1415.

[46] Xu ZX, Yin XX, Sievenpiper DF. Adiabatic mode-matching techniques for coupling between conventional microwave transmission lines and onedimensional impedance interface waveguides. Phys. Rev. Appl. 2019, 11, 044071.

[47] Muench JE, Ruocco A, Giambra MA, et al. Waveguide-integrated, plasmonic enhanced graphene photodetectors. Nano Lett. 2019, 19, 7632–7644.

[48] Kim S, Nah J. Jo I, et al. Realization of a high mobility dual-gated graphene field-effect transistor with $Al_2O_3$ dielectric. Appl. Phys. Lett. 2007, 94, 062107.

[49] Born M, Wolf E. Elements of the theory of diffraction, in Principles of Optics, 7th ed. Cambridge University Press Cambridge, 2005.

[50] Ma Q, Gabor NM, Andersen TI, et al. Competing channels for hot-electron cooling in graphene. Phys. Rev. Lett. 2014, 112, 1–5.

[51] Dollfus P, Nguyen VH, Saint-Martin J. Thermoelectric effects in graphene nanostructures. J. Phys. Condens. Matter. 2015, 27, 133204.

[52] Viti L, Hu J, Coquillat D, Politano A, Knap W, Vitiello MS. Efficient terahertz detection in black-phosphorus nano-transistors with selective and controllable plasma-wave, bolometric and thermoelectric response. Sci. Rep. 2016, 6, 20474.

[53] Song JCW, Rudner MS, Marcus CM, Levitov LS. Hot carrier transport and photocurrent response in graphene. Nano Lett. 2011, 11, 4688–4692.

[54] Avtech AVR series, medium to high voltage general purpose pulse generators (Accessed March 2020, www.Avtechpulse.Com/Catalog/Avr-1-2-3-4_rev17.Pdf.).

[55] Lisauskas A, Ikamas K, Massabeau S, et al. Field-effect transistors as electrically controllable nonlinear rectifiers for the characterization of terahertz pulses. APL Photonics 2018, 3, 051705.

[56] Das A, Pisana S, Chakraborty B, et al. Monitoring dopants by Raman scattering in an electrochemically top-gated graphene transistor. Nat. Nanotechnol. 2008, 3, 210–215.

[57] Basko DM, Piscanec S, Ferrari AC. Electron-electron interactions and doping dependence of the two-phonon Raman intensity in graphene. Phys. Rev. B 2009, 80, 1–10.

[58] Cançado LG, Jorio A, Martins Ferreira EH, et al. Quantifying defects in graphene via Raman spectroscopy at different excitation energies. Nano Lett. 2011, 11, 3190–3196.

[59] Neumann C, Reichardt S, Venezuela P, et al. Raman spectroscopy as probe of nanometre-scale strain variations in graphene. Nat. Commun. 2015, 6, 1–7.

[60] Vitiello MS, Scamarcio G. Measurement of subband electronic temperatures and population inversion in THz quantum-cascade lasers. Appl. Phys. Lett. 2005, 86, 111115.